\documentclass[preprint, prX]{revtex4}

\usepackage{amsmath}    
\usepackage{graphicx}   
\usepackage{verbatim}   
\usepackage{color}      
\usepackage{subfigure}  
\usepackage{hyperref}   
\usepackage{amssymb}    
\usepackage{epsfig}
\usepackage{graphics,graphicx}
\usepackage{setspace}
\usepackage{url}
\usepackage{algorithm,algorithmic}

\begin{document}

\begin{abstract}
Measuring time in mass sporting competitions is unthinkable manually today because of their long duration and unreliability. Besides, automatic timing devices based on the RFID technology have become cheaper. However, these devices cannot operate stand-alone. To work efficiently, they need a computer timing system for monitoring results. Such system should be capable of processing the incoming events, encoding and assigning results to a individual competitor, sorting results according to the achieved time and printing them. In this paper, a domain-specific language named EasyTime will be defined. It enables controlling an agent by writing events to a database. Using the agent, the number of measuring devices can be reduced. Also, EasyTime is of a universal type that can be applied to many different sporting competitions

\textit{To cite paper as follows: I. Fister Jr., I. Fister, Measuring Time in Sporting Competitions with the Domain-Specific Language EasyTime, Elektrotehni\v{s}ki vestnik, 78(1-2): 36-41, 2011
}

\end{abstract}

\title{Measuring Time in Sporting Competitions with the Domain-Specific Language EasyTime}

\author{Iztok Fister Jr.}
\altaffiliation{University of Maribor, Faculty of electrical engineering and computer science
Smetanova 17, 2000 Maribor}
\email{iztok.fister@guest.arnes.si}

\author{Iztok Fister}
\altaffiliation{University of Maribor, Faculty of electrical engineering and computer science
Smetanova 17, 2000 Maribor}
\email{iztok.fister@uni-mb.si}

\maketitle

\section{Introduction}

Not long ago, time in sporting competitions was measured manually by timekeepers. The measured time was assigned to the starting number of competitors who were arranged according to the final result and the competitor's category. With the arrival of the Radio Frequency Identification technology (RFID)~\cite{web:RFID2010}, the cost of the measurement technology (like ChampionChip~\cite{web:ChampionChip2010} and RFID Race Timing System~\cite{web:RFID2010}) was reduced. It thus became accessible to a wider class of users, e.g. sport clubs, organizers of sporting competitions etc. They then began to compete with the existing monopolies (Timing Ljubljana~\cite{web:Timing2010}) by measuring results in smaller sporting competitions.

Besides the measuring technology in a sporting competition, a flexible computer timer system is also needed allowing measuring of various sporting competitions with an arbitrary number of measuring places, online time tracking, printing result lists and ensuring reliability and security. Flexibility of such a system can be increased by using the domain-specific language (DSL) EasyTime.

The domain-specific languages\cite{Mernik:2005} are suitable for the application domain and have definite advantages over general-purpose languages in a specific domain. First of all, these advantages are expressed in a higher expressive power and, therefore, higher pro\-duc\-ti\-vi\-ty, ease of use (even for domain experts that are not programmers), easier verification and optimization. EasyTime is used to configure agents to write an event that arises in a measuring device into a database. Thus, agents are crucial elements of the timing system. When carefully configured, the number of the necessary devices can be decreased.

The structure of the rest of the paper is as follows. In Section~2, problems of time measuring in sporting competitions are described. The focus will primarily be on the triathlon competitions which are the har\-dest to be efficiently measured because of the three different disciplines involved. In Section 3, EasyTime is presented in detail. Section 4 describes performance of a program written in EasyTime. The paper ends with a short analysis of the performed work and plans for the future work.

\section{Measuring Time in Sporting Competitions}

In practice, the measuring time in sporting competitions can be measured $manually$ (classically or with a computer timer) or $automatically$ (by using a measuring device). The computer timer is an application that runs usually on a workstation (portable computer) and measures the real time. Through this, a processor tact is exploited. The processor tact is the velocity with which the processor executes computer instructions. The computer timer enables tracking events that are triggered by a competitor crossing over the measuring place (MP) similarly to a measuring device. However, in that case, the event is triggered by an operator on the computer pressing the appropriate key on the keyboard. The operator generates events in a form of triples $\langle \#,MP,TIME \rangle$, where $\#$ denotes the starting number of the competitor, $MP$ the measuring place at which the event takes places, and $TIME$ the timestamp that is generated by the device at the moment of triggering and represents the number of seconds since 1.1.1970 at 0:0:0.

Today, the measuring devices are usually based on the RFID technology~\cite{web:RFID2010}. Thus, identification is performed by electro-magnetic wave motions within range of radio frequencies that are radiated by antenna fields. The measuring devices consist of:
\begin{itemize}
  \item RFID tag readers,
  \item primary memory,
  \item LCD display, and
  \item numeric keyboard.
\end{itemize}
Here can be several antenna fields connected to the device. These fields represent a particular measuring place. Competitors trigger events by crossing over the antenna field with passive RFID tags that bear their identification numbers. These numbers are unique and different from the starting numbers. The event on the measuring device is represented in a quadruple form $\langle \#,RFID,MP,TIME \rangle$, where the RFID identification number is also added to the triplet. The accuracy of those measuring devices is usually limited to $1/10$ second that is enough for the propositions of the referee associations.

The measuring devices and workstations with an installed computer timer that represent the measuring places in the timing system can be connected to a local-area network (LAN). With these devices we communicate via a control program, i.e. an agent that runs on a database server. The agent gets connected with the measuring device via a suitable TCP/IP socket that supports an appropriate TCP/IP protocol. The measuring devices usually support the protocol $Telnet$ that is easy to implement and enables a text stream-oriented communication. The agent communicates with the manual timer via a file transfer protocol.

\subsection{An Example: Measuring Time in a Triathlon}

Special requirements appear in triathlon competitions where there are three disciplines to be dealt with in one competition. Therefore, we will focus on that problem in this paper.

The first triathlon competition was performed in the USA in 1975. The competition is regarded today as an olympic discipline in which the competitor starts with swimming, then rides a bicycle and fi\-ni\-shes with running. All the three activities are performed sequentially and continuously. For the summary time, delays in both transitions are added. In the first transition, the competitor goes from swimming to bicycling, while in the second transition, he/she moves from bicycling to running. Today, there are many kinds of triathlon competitions. They are distinguished according to the length of particular courses. Normally, organizers use circular courses of shorter lengths (laps) over which competitors need to pass multiple times. However, this makes measuring considerably more difficult, since the number of laps also needs to be counted.

\begin{figure*}[htb]  
\vspace{-5mm}
    \begin{center}
        \includegraphics {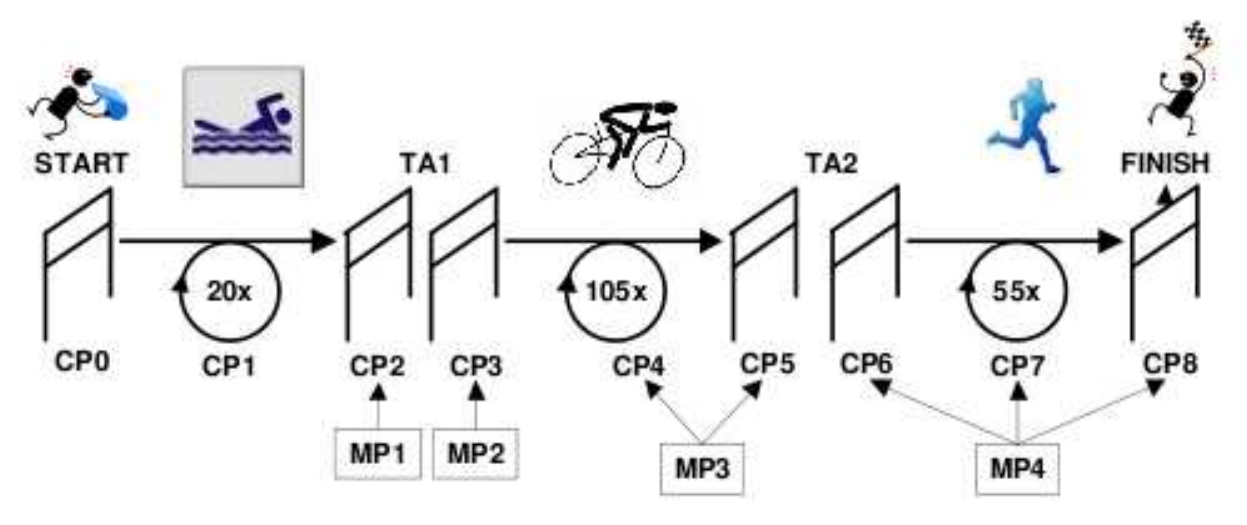}  
        \caption{Definition of control points in the triathlon competition}
        \label{pic:slika_1}
    \end{center}
\vspace{-5mm}
\end{figure*}

As seen from Fig.~\ref{pic:slika_1}, the measuring time in triathlon competitions is divided into nine control points (CP). Control points are locations on the course where organizers need to track the measured time that can be $intermediate$ or $finish$. In Fig.~\ref{pic:slika_1}, we are dealing with a double ultra triathlon (7.6 kilometers of swimming, 360 kilometers of bicycling and 84 kilometers of running), where the length of the swimming course is 380 meters (or 20 laps), the bicycle course is 3.4 kilometers (or 105 laps) and the running course is 1.5 kilometers (or 55 laps).

The summary time of a triathlon competition consists of five final times (the swimming time SWIM (CP2), the first transition time TA1 (CP3), the bicycling time BIKE (CP5), the second transition time TA2 (CP6) and running time RUN (CP8)) as well as three intermediate times (the intermediate swimming time (CP1), the intermediate bicycling time (CP4) and the intermediate running time (CP7)). With the intermediate times, the number of laps ROUND\_$x$ and achieved time INTER\_$x$ are measured. Here, $x={1,2,3}$ denotes a particular course.

Suppose that a measuring device with two measuring places (MP3 and MP4) is available for the measurement in Fig.~\ref{pic:slika_1} and the competition is performed at one location. In such case, the last crossing over the MP3 can indicate the CP5 time, the first crossing over the MP4 the CP7 time and the last crossing over the MP4 the final result (CP8). The measuring places MP1 and MP2 are measured manually by the computer timer. Finally, the number of measuring points can be reduced by three if the timing system and the control points are appropriately set up. Thus, 162 events for each competitor can be measured with one measuring device (or 87.5\%). Moreover, the measurement technology for measuring swimming in sees and lakes still being expensive and thus still being measured by referees manually, almost 98\% of all events at such competition can be measured.

\section{DSL EasyTime}

With EasyTime, various measurements need to be described. This is made with the timing system. Moreover, reduction in the number of measuring devices is expected when using more complex measuring time. EasyTime enables describing the rules for controlling the agent before the event registering at a measuring place is recorded into the database. However, each program written in EasyTime needs to be compiled before an execution. Compiling consists of:
\begin{itemize}
  \item syntax analysis and
  \item code generation.
\end{itemize}
Syntax analysis is performed by a syntax analyzer (also parser) that the program written in EasyTime compiles into an intermediate code. From this code, a code-generator generates an executable code for a virtual machine and stores it in a database. The database table in which the code is stored is named the rule table because this code contains rules for controlling the agents. In the rest of the paper, the characteristics of the parser and code generator are presented in detail.

\subsection{The Parser}

\begin{figure*}[!htb]
\vspace{-5mm}
    \begin{center}
        \includegraphics [scale=0.7] {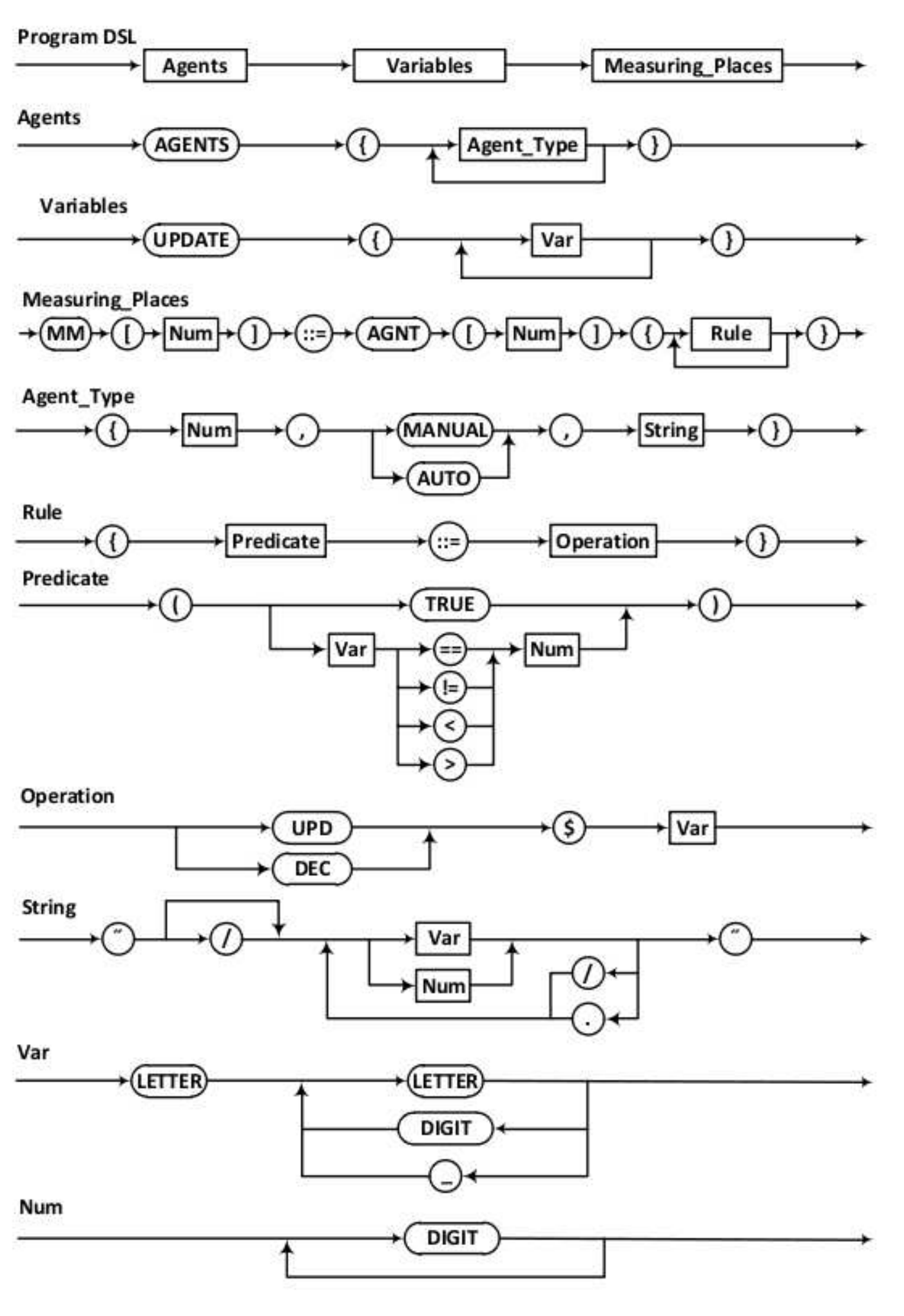}
        \caption{Sintax of DSL EasyTime}
        \label{pic:slika_5}
    \end{center}
\vspace{-5mm}
\end{figure*}

Before developing a parser, a syntax for EasyTime needs to be defined. The syntax is a set of rules in which the structure of correct statements (grammar) is defined~\cite{Aho:2007}. The syntax of EasyTime is presented in syntax diagrams in Fig.~\ref{pic:slika_5}. The syntax diagrams are the most suitable form for writing parsers. It has the same expressive power as the BNF (Backus Naur Form) notation~\cite{Aho:1972a}, i.e. a notation technique suitable for context-free grammars. The parser was developed in the programming language C/C++~\cite{Wirth:1978}.

\begin{table}[!htb]
\caption{Names of variables in the table $\mathit{RESULTS}$}
\label{tab:tab2}
\smallskip
\small
\begin{center}
\begin{tabular}{ | l | l |}
\hline
  \textbf{Variable} & \textbf{Description} \\
\hline
  ROUND\_1 & Number of laps of swimming \\
  INTER\_1 & Intermediate time 1 (CP1) \\
  SWIM & Finish time of swimming (CP2) \\
  TRANS\_1 & Transition time 1 (CP3) \\
  ROUND\_2 & Number of laps of bicycling \\
  INTER\_2 & Intermediate time 2 (CP4) \\
  BIKE & Finish time of bicycling (CP5) \\
  TRANS\_2 & Transition time 2 (CP6)\\
  ROUND\_3 & Number of laps of running \\
  INTER\_3 & Intermediate time 3 (CP7) \\
  RUN & Final time of triathlon (CP8) \\
\hline
\end{tabular}
\end{center}
\normalsize
\end{table}

The EasyTime program consists of definitions of:
\begin{itemize}
  \item agents,
  \item variables, and
  \item measuring places.
\end{itemize}

The definition of agents described with the statement $Agents$ is represented in the syntactic diagram with the same name in Fig.~\ref{pic:slika_5}. Variables denoting the names of columns in the database (RESULTS table) and representing the control points can be seen in Table~\ref{tab:tab2}. However, these needs to be defined before using them in the $Measuring-Places$ statement that defines the valid rules for a particular measuring place. The rules are in the form of $\langle \textnormal{Predicate}\rangle$::=$\langle \textnormal{Operation} \rangle$. Note that in EasyTime character $\$$ is set before the name of a variable.

For example, the definition of agents in EasyTime describes the measuring time in Fig.~\ref{pic:slika_1} as presented in Program~\ref{alg:agents}, where the first agent saves the results of the manually measured time in the directory $/home/DC2$ and the second agent automatically obtains data from the measuring device with the IP address $192.168.225.100$ via the UDP protocol on port 9999.

\begin{algorithm}[!htb]
\caption{Definition of agents}
\label{alg:agents}
\begin{algorithmic}[1]
\STATE AGENTS \{
\STATE \ \ \ \ \{1,MANUAL,\textquotedblleft /home/DC2/res.ets\textquotedblright\}
\STATE \ \ \ \ \{2,AUTO,\textquotedblleft 192.168.225.100/UDP/9999\textquotedblright\} \}
\end{algorithmic}
\end{algorithm}

\begin{algorithm}[!htb]
\caption{Definition of measuring places}
\label{alg:mm}
\begin{algorithmic}[1]
\STATE MM[1] ::= AGNT[1] \{
\STATE \ \ \ \ \{ (TRUE) ::= UPD \$SWIM \}
\STATE \ \ \ \ \{ (TRUE) ::= DEC \$ROUND\_1 \} \}
\STATE MM[2] ::= AGNT[1] \{
\STATE \ \ \ \ \{ (TRUE) ::= UPD \$TRANS\_1 \} \}
\STATE MM[3] ::= AGNT[2] \{
\STATE \ \ \ \ \{ (TRUE) ::= UPD \$INTER\_2 \}
\STATE \ \ \ \ \{ (TRUE) ::= DEC \$ROUND\_2 \}
\STATE \ \ \ \ \{ (ROUND\_2 == 0) ::= UPD \$BIKE \} \}
\STATE MM[4] ::= AGNT[2] \{
\STATE \ \ \ \ \{ (TRUE) ::= UPD \$INTER\_3 \}
\STATE \ \ \ \ \{ (ROUND\_3 == 55) ::= UPD \$TRANS\_2 \}\}
\STATE \ \ \ \ \{ (TRUE) ::= DEC \$ROUND\_3 \}
\STATE \ \ \ \ \{ (ROUND\_3 == 0) ::= UPD \$RUN \} \}
\end{algorithmic}
\end{algorithm}

The rules for measuring places given in Fig.~\ref{pic:slika_1} are determined in EasyTime with the source code presented in Program~\ref{tab:tab4}. Each measuring place is denoted with its identification number and connected with an appropriate agent. The rules for the measuring place 4, for example, determine that the event generated by a measuring device at first updates the intermediate time of running INTER\_3. In such case, the competitor crosses over the measuring place for the first time (predicate ROUND\_3==55) and the time of the second transition TRANS\_2 is updated. Then, a decrementing number of laps ROUND\_3 follows. Finally, the agent announces the final result when the competitor is in the last lap (predicate ROUND\_3==0) and, obviously, the variable RUN is updated.

\subsection{The Code Generator}

The code generator~\cite{Aho:1972b} is performed if the syntax analysis has been successfully completed. When the program fails, the parser provides error messages and stops. The code generator generates the code for each measuring place separately. The generated code is saved in the database. The code generator is developed in the programming language C/C++ as well.

\begin{figure}[!htb]
    \begin{center}
        \includegraphics[scale=0.8]{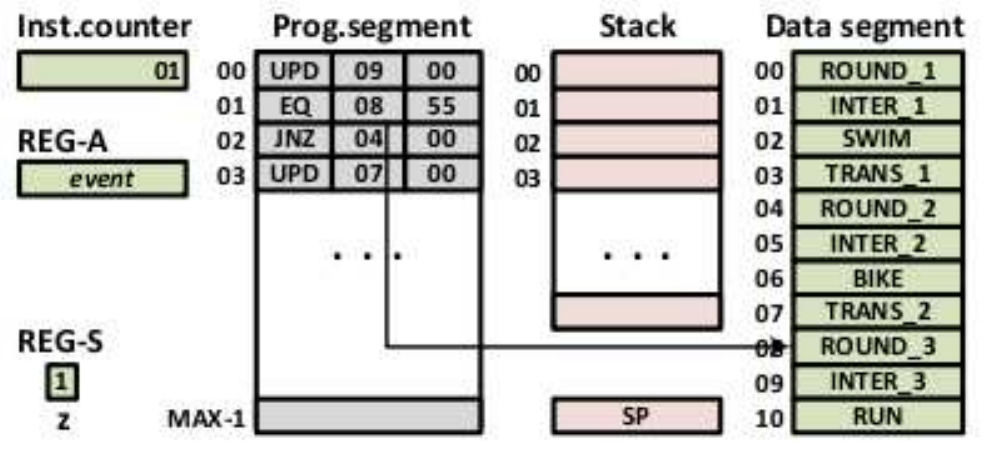}
        \caption{Architecture of the virtual machine}
        \label{pic:slika_7}
    \end{center}
\end{figure}

The architecture of the process for which the code is generated needs to be defined before generation takes place. The compiled EasyTime program is executed on a virtual machine. In line with the process parallelization (multi-threading), the virtual machine is for each measuring place defined separately. The architecture of the virtual machine (Fig.~\ref{pic:slika_7}) is simple. It only consists of a program segment, stack, data segment, instruction counter, and program and status registers. The generated code is loaded into the program segment. On the stack, arithmetical-logical operations are executed. The data segment consists of variables from the database. The instruction counter points to the instruction in the program segment that is currently interpreted. The data register (REG-A) holds the timestamp of the event that is treated by the agent. The status register (REG-S) holds the status of predicates in a binary form ($TRUE$ if $z==1$ or $FALSE$ if $z==0$).

The program instructions loaded into the program segment of a particular virtual machine from the database are defined as a triple $\langle op,p1,p2 \rangle$, where $op$ denotes an operation code, and $p1$ and $p2$ are parameters (variable or constant). An instruction set consists of logical instructions EQ and NEQ, ope\-ra\-tions UPD, DEC and STOP, and branch instruction JNZ. The logical instructions affect the setup of the status register that controls interpretation of the branch instruction.

The code generator generates the code from the data structures that are built by the parser. In practice, the code is only generated from a data structure built from the $Measuring\_Places$ statement. The names of variables assembled in the table are translated into their addresses in the data segment. The addresses appear in instructions as parameters. The data structure built from the $Agents$ statement is designed for the configuration of a particular agent before its execution. An example of a code generated (in a symbolic form) from the source code (Program~\ref{alg:agents}) determining the rules of the agent controlling MP3 is presented in Table~\ref{tab:tab4}.

\begin{table}[!htb]
\caption{The program code for the measuring place 3}
\label{tab:tab4}
\smallskip
\small
\begin{center}
\begin{tabular}{  l  l  l  l }
\hline
  \textbf{IC} & \textbf{OP.C.} & \textbf{P1} & \textbf{P2} \\
\hline
  00 & UPD  & 5 & 0 \\
  01 & DEC  & 4 & 0 \\
  02 & EQ   & 4 & 0 \\
  03 & JNZ  & 5 & 0 \\
  04 & UPD  & 6 & 0 \\
  05 & STOP & 0 & 0 \\
\hline
\end{tabular}
\end{center}
\normalsize
\end{table}

From the example given in Table~\ref{tab:tab4}, it can be seen that the generated code is optimized because from three lines of a source code six lines of the executable code are obtained. As each instruction recorded in the database is four bytes long (the operation code, two parameters and delimiter ';'), the size of the compiled program is not critical for the database.

\section{Operation of the Agent}

\begin{figure}[!htb]
\vspace{-5mm}
    \begin{center}
        \includegraphics [scale=1.0]{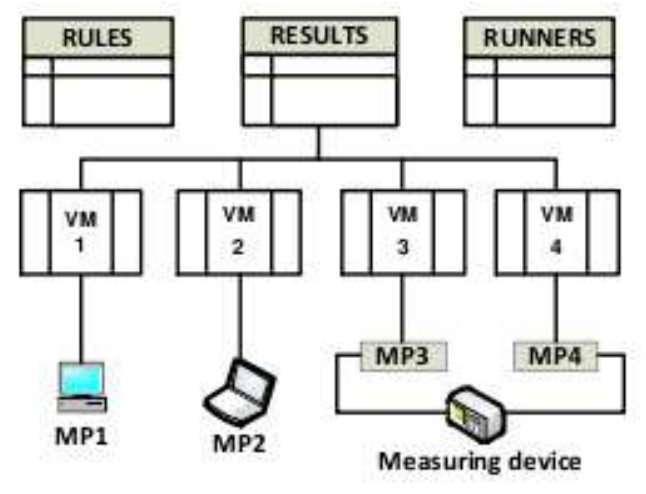}      %
        \caption{Executing environment of the EasyTime program}
        \label{pic:slika_3}
    \end{center}
\vspace{-5mm}
\end{figure}

The agent that is controlled with the EasyTime program can process the following events:
\begin{itemize}
  \item batch: manual mode of operation ($\textit{MANUAL}$),
  \item online: automatic mode of operation ($\textit{AUTO}$).
\end{itemize}
In batch processing, events assembled in a text file are read and written in an appropriate database by the agent. Typically, events captured with computer timers are batch-processed. In this processing mode, the agent checks every second if the file configured in the agent table exists or not. In case that it exists, batch processing begins. At the end of processing, the file is archived and then deleted. Online processing is event-oriented, i.e. each event registered by the measuring device is processed in time. The executing environment introduced by the compiled program written in EasyTime for the measuring time in the triathlon competition as illustrated in Fig.~\ref{pic:slika_1}, is presented in Fig.~\ref{pic:slika_3}.

In both processing modes, the agent operates with the following database tables: rules ($\mathit{RULES}$), competitors ($\mathit{RUNNERS}$) and results ($\mathit{RESULTS}$). When the agent starts, initialization of the virtual machine for each measuring place is executed. Initialization consists of loading the program code from the rule database table. The code is loaded only once. At the same time, variables in the data segment are initialized. Recording the event processed by the agent can be divided into the following phases:
\begin{itemize}
  \item reconstruction of the event: the competitor is identified either over the starting number (\#) of $\mathit{RFID}$ tag (Fig.~\ref{pic:slika_3}), while the appropriate virtual machine is determined according to the measuring place $\mathit{MP}$ and the data register is loaded by the timestamp $\mathit{TIME}$,
  \item reading of the results: from the database table $\mathit{RESULTS}$, the current results of the competitor triggering the event are read,
  \item mapping of the results: the current results of the competitor are mapped to the data segment of a particular virtual machine,
  \item code interpretation: the instruction counter is set to zero and the program that was loaded into the program segment starts to execute, and
  \item recording of the results: the results from the data segment of a particular virtual machine are written to the database table $\mathit{RESULTS}$.
\end{itemize}
The program in a particular virtual machine is interpreted sequentially, i.e. instruction by instruction, until the instruction STOP is detected.

\section{Conclusion}

When developing universal software for the measuring time in sporting competitions the problem of flexibility in the timing system is often encountered. To cope with the issue, the domain specific language EasyTime was developed enabling rapid adaptation of a timing system to the demands of various sporting competitions. To monitor a new competition, modifications to the source program written in EasyTime need to be compiled and the agent needs to be restarted. As a result, the agent is ready to run in a completely new environment. Using EasyTime in practice shows that organizers can no longer employ specialized and expensive companies to measure time in sporting competitions. However, for large sporting competitions, configuration of the timing system can be simplified. Our future work will be towards upgrading EasyTime with a domain-specific modeling language further simplify the configuration process of the timing system.

\bigskip{\small \smallskip\noindent Updated 25 November 2012.}

\begin{thebibliography}{99}
\bibitem{Mernik:2005} M. Mernik, J. Heering, A. Sloane: When and how to develop domain-specific languages, 2005, ACM computing surveys, vol. 37, no. 4, pp. 316-344
\bibitem{Finkenzeller:2010} K. Finkenzeller: RFID Handbook, 2010, John Willey, Chichester, UK
\bibitem{Wirth:1978} N. Wirth: Algorithms + Data Structures = Programs, 1978, Prentice Hall PTR, Upper Saddle River, NJ, USA
\bibitem{Aho:1972a} A.V. Aho, J.D. Ullman: The theory of parsing, translation, and compiling (Volume I: Parsing), 1972, Prentice Hall PTR, Upper Saddle River, NJ, USA
\bibitem{Aho:1972b} A.V. Aho, J.D. Ullman: The theory of parsing, translation, and compiling (Volume II: Compiling), 1972, Prentice Hall PTR, Upper Saddle River, NJ, USA
\bibitem{Aho:2007} A.V. Aho, M.S. Lam, R. Sethi, J.D. Ullman: Compilers: Principles, Techniques, and Tools with Gradiance, 2007, Prentice Hall PTR, Upper Saddle River, NJ, USA
\bibitem{web:ChampionChip2010} ChampionChip2010, http://www.championchip.com
\bibitem{web:RFID2010} RFIDTechnology2010, http://www.rfidtiming.com
\bibitem{web:Timing2010} Timing2010, http://www.timingljubljana.si
\end{thebibliography}
\end{document}